\documentclass[fleqn,usenatbib]{mnras}

\usepackage{newtxtext,newtxmath}

\usepackage[T1]{fontenc}
\usepackage{ae,aecompl}

\usepackage{enumerate}
\usepackage[normalem]{ulem}
\usepackage[usenames]{color}

\usepackage{graphicx}	
\usepackage{amsmath}	
\usepackage{amssymb}	

\newcommand{\changaMM}{{\sevensize MANGA}}
\newcommand{\changa}{{\sevensize CHANGA}}
\newcommand{\charm}{{\sevensize CHARM++}}
\newcommand{\mesa}{{\sevensize MESA}}
\newcommand{\be}{\begin{eqnarray}}
\newcommand{\ee}{\end{eqnarray}}
\newcommand{\rsun}{\ensuremath{\textrm{R}_{\odot}}}
\newcommand{\msun}{\ensuremath{\textrm{M}_{\odot}}}

\newcommand{\grad}{\ensuremath{\boldsymbol{\nabla}}}
\newcommand{\vel}{\ensuremath{\boldsymbol{v}}}
\newcommand{\velCM}{\ensuremath{\boldsymbol{v}_{\rm CM}}}
\newcommand{\velp}{\ensuremath{\boldsymbol{v}'}}

\newcommand{\ddt}[1]{\ensuremath{\frac{\partial #1}{\partial t}}}

\newcommand{\state}{\ensuremath{\boldsymbol{\mathcal{U}}}}
\newcommand{\charge}{\ensuremath{\boldsymbol{U}}}

\newcommand{\flux}{\ensuremath{\boldsymbol{\mathcal{F}}}}
\newcommand{\fluxV}{\ensuremath{\boldsymbol{F}}}
\newcommand{\source}{\ensuremath{\boldsymbol{\mathcal{S}}}}
\newcommand{\sourceV}{\ensuremath{\boldsymbol{S}}}

\newcommand{\normal}{\ensuremath{\hat{\boldsymbol{n}}}}

\newcommand{\meshv}{\ensuremath{\boldsymbol{w}}}
\newcommand{\facev}{\ensuremath{\boldsymbol{\tilde{w}}_{ij}}}
\newcommand{\facer}{\ensuremath{\boldsymbol{\tilde{r}}_{ij}}}

\newcommand{\dt}{\ensuremath{\delta t}}

\newcommand{\Etoti}{\ensuremath{E_{\rm tot,i}}}

\newcommand{\mcore}{\ensuremath{m_{\rm c,1}}}
\newcommand{\mcomp}{\ensuremath{m_{\rm c,2}}}
\newcommand{\mbound}{\ensuremath{m_{\rm g,b}}}
\newcommand{\vbound}{\ensuremath{\boldsymbol{v}_{\rm g,b}}}
\newcommand{\munbound}{\ensuremath{m_{\rm g,u}}}
\newcommand{\funbound}{\ensuremath{f_{\rm unb}}}
\newcommand{\vcore}{\ensuremath{\boldsymbol{v}_{1}}}
\newcommand{\vcomp}{\ensuremath{\boldsymbol{v}_{2}}}
\newcommand{\rapo}{\ensuremath{r_{\rm apo}}}
\newcommand{\rperi}{\ensuremath{r_{\rm peri}}}

\newcommand{\lp}[1]{\textrm{\color{red} #1}}
\newcommand{\pc}[1]{\textrm{\color{red} #1}}

\renewcommand{\sout}[1]{}
\renewcommand{\lp}[1]{#1}
\renewcommand{\pc}[1]{#1}

\title[Common Envelope Evolution on a Moving Mesh]{Common Envelope Evolution on a Moving Mesh}

\author[]{ Logan J. Prust$^1$\thanks{LJP: ljprust@uwm.edu, PC: chang65@uwm.edu, } and Philip Chang$^{1,2}$
\\
$^1$ Department of Physics, University of Wisconsin-Milwaukee, 3135 North Maryland Avenue, Milwaukee, WI 53211, USA\\
$^2$ Center for Computational Astrophysics, Flatiron Institute, 162 Fifth Avenue, New York, NY, 10010, USA
}

\date{Accepted XXX. Received YYY; in original form ZZZ}

\pubyear{2019}

\begin{document}
\label{firstpage}
\pagerange{\pageref{firstpage}--\pageref{lastpage}}
\maketitle

\begin{abstract}
We outline the methodology of simulating common envelope evolution (CEE) with the moving-mesh code \changaMM.  We extend \changaMM\ to include multiple time-steps. This provides substantial speedups for problems with large dynamic range.  We describe the implementation of realistic equations of state relevant in stellar structure and the generation of suitable initial conditions.  We then carry out two example simulations of a 2 \msun\ red giant with a 0.36 \msun\ core and a 1 \msun\ companion undergoing CEE for 240 days.  In one simulation the red giant is set into corotation with the orbital motion and in the other it is non-rotating.  We find that the separation between the companion and red giant core shrinks from 52 \rsun\ to 3.6 \rsun\ and 3.2 \rsun\ respectively, ending with an eccentricity of 0.1. We also find that 66 and 63 per cent of the envelope mass is ejected. This is higher than in many previous works.  Several reasons for this are discussed. These include our inclusion of recombination energy.  Our simulations show that putting giants in corotation increases the fraction of mass ejected from the system and results in a larger final orbital separation.  We conclude that the entire envelope of the red giant might be ejected during the plunge phase of CEE in this region of parameter space.
\end{abstract}




\section{Introduction}

Common envelope evolution (CEE) is the name given to a brief but important phase in the evolution of a binary system where two stars -- one compact, one giant -- share a common envelope \citep[for a review see][]{2013A&ARv..21...59I}.  As the core of the giant and the compact star orbit each other, loss of orbital energy and angular momentum drive the two stellar centres close to one another and eject the envelope from the system.  CEE is a critical process in the lives of these binary systems and is responsible for progenitors of Type Ia supernova (potentially), X-ray binaries, double white dwarfs, double neutron stars and possibly the merging double black hole and double neutron star systems recently discovered by Advanced LIGO \citep{2013A&ARv..21...59I,2016Natur.534..512B}.


In spite of its importance, CEE is not well understood.  Our knowledge of CEE mainly comes from population necessity rather than theory or direct observation.  Part of this problem is the \lp{\sout{number of different} diverse} physics, including hydrodynamics, convection, turbulence, accretion, nuclear burning, radiation, self-gravity, ionization/recombination, jets and magnetism \lp{involved}.  In addition, the physical time-scales span the dynamical time of compact remnants or cores (milliseconds to seconds) to the many thermal times of an envelope (years) \citep{2008ASSL..352..233W,2013A&ARv..21...59I}.

In the presence of the different physics and time-scales, astronomers have mainly employed conserved quantities -- energy and angular momentum -- to roughly model CEE.  These are known as the energy formalism \citep{1984ApJ...277..355W,1988ApJ...329..764L} and the angular momentum (or $\gamma$) formalism \citep{2000A&A...360.1011N,2005MNRAS.356..753N}.  
To make progress beyond these formalisms, fully 3-D numerical tools must be brought to bear.
Here the defining numerical studies are those carried out by \citet{2012ApJ...744...52P}, \citet{2012ApJ...746...74R}, \cite{2015MNRAS.450L..39N} and \citet{	  2016ApJ...816L...9O}.  These simulations generally find a wide variation in efficiencies from about 5 to 100 per cent. \citet{2015MNRAS.450L..39N} use a smooth particle hydrodynamics (SPH) code, \citet{2012ApJ...744...52P} used both a SPH and an Eulerian code, \citet{2012ApJ...746...74R} used an adaptive mesh refinement Eulerian code and \citet{	  2016ApJ...816L...9O} used perhaps the most interesting code of all -- the moving-mesh (MM) code, {\sevensize AREPO}.

{\sevensize AREPO} is a class of arbitrary Lagrangian-Eulerian (ALE) or MM schemes that have been devised as an effort to capture the best characteristics of both Lagrangian and Eulerian approaches, combining superior conservation properties of Lagrangian schemes with the superior shock capturing of Eulerian schemes.
\citet[][hereinafter S10]{2010MNRAS.401..791S} described a usable ALE scheme that has proven successful and is implemented in the code {\sevensize AREPO}. The scheme relies on a Voronoi tessellation \citep{1992stca.book.....O} to generate a well-defined, unique, and continuously deformable mesh for an arbitrary distribution of points.

S10 argued that the use of ALE schemes is important to maintain the Galilean invariance of Eulerian schemes.  It has also been argued that these schemes are superior at capturing boundary layer instabilities such as Kelvin-Helmholtz instabilities to SPH and Eulerian grid schemes (S10, but also see \citealt{	  2016mnras.455.4274l}).  In any case, MM methods do seem ideal to model colliding galaxies or stars.  In particular, {\sevensize AREPO} has been used in a number of different problems including cosmological galaxy formation \citep[see for instance][]{2014MNRAS.444.1518V}, and stellar mergers \citep{2015ApJ...806L...1Z,2016ApJ...816L...9O}.

Recently, we have developed a moving-mesh hydrodynamic solver for \lp{the N-body simulation code} \changa\ \lp{(Charm N-body GrAvity solver) \citep{jetley2008,jetley2010,2015comac...2....1m}.} \lp{\sout{which we call} We call this moving-mesh solver} \changaMM\ \citep[][hereinafter C17]{Chang+17}.  The solver is based on the scheme described by S10, but also utilizes advances in gradient estimation \citep{2016MNRAS.459.1596S} and limiters \citep{2011ApJS..197...15D}.  We also use an alternative method to construct the Voronoi tessellation (\citealt{2009Chaos..19d1111R}; C17).  More recently, we have added magnetohydrodynamics (Chang, in preparation) and radiation hydrodynamics (Chang, Davis, \& Jiang in preparation) to \changaMM.

\pc{\sout In addition,} \changa\ \pc{\sout{is unique among astrophysical codes in that it }}uses the \charm\ language and run-time system \citep{KaleKrishnan96} for parallelization rather than a custom message-passing interface design.\footnote{Other astrophysical codes that use \charm\ include {\sevensize ENZO-P} and {\sevensize FVMHD3D}\ \citep{2012ApJ...758..103G}.}  The use of \charm\ promises that \changa\ will be much more scalable than previous astrophysical codes. In particular, \changa\ has demonstrated strong scaling on single time-stepping problems with $12$ billion particles to 512K cores (with 93\% efficiency) and on multi-time-stepping problems with 52 million particles to 128K cores \citep{2015comac...2....1m}.

Given the importance of CEE and the recent advances in numerical computation, we apply \changaMM\ to study CEE using MM techniques.  In this initial work, we describe our methodology, highlighting the algorithmic improvements to \changaMM\ to enable this study.  In particular, we describe the implementation of individual time-steps, the incorporation of a realistic equation of state and the generation of realistic stars and initial conditions.  We also show one example CEE evolution computed both with and without the giant set into corotation with the binary orbit. 

We have organized this initial paper as follows.  In section \ref{sec:numerical}, we summarize the algorithm of \changaMM, highlighting improvements made since C17. We also clarify the half time-step prediction that was not well described by C17. We describe the implementation of multistepping in section \ref{sec:multistep} and realistic equations of state in section \ref{sec:eos}.  We describe how we use the stellar evolution code \mesa\ to produce 1-D models of red giants that we map into 3-D unstructured grids in section \ref{sec:ICs} and use these models as initial conditions for CEE.  In section~\ref{sec:results}, we describe the results of our simulations of a 2 \msun\ red giant and a 1 \msun\ companion.  We discuss our results and close in section \ref{sec:discussion}. 

\section{Methodology}


\subsection{Numerical Approach}\label{sec:numerical}

We begin with a brief summary of \changaMM.  This is mainly covered by C17 but the algorithm has been modified since C17 which we highlight below. \changaMM\ solves the the Euler equations\pc{, e.g., the conservation of mass equation, conservation of momentum and conservation of energy}, which written in conservative form are
\be
\ddt{\rho} + \grad\cdot\rho\vel &=& 0 \label{eq:continuity},\\
\ddt{\rho\vel} + \grad\cdot\rho\vel\vel + \grad P &=&-\rho\grad\Phi\label{eq:momentum}
\ee
\begin{flushleft}
and
\end{flushleft}
\be
\ddt{\rho e} + \grad\cdot\left(\rho e + P\right)\vel &=& -\rho\vel\cdot\grad\Phi\label{eq:energy},
\ee
where $\rho$ is the density, $\vel$ is the velocity, $\Phi$ is the gravitational potential, $e= \epsilon + v^2/2$ is the specific energy, $\epsilon$ is the internal energy, and $P(\rho, \epsilon)$ is the pressure.  Equations (\ref{eq:continuity}) to (\ref{eq:energy}) can be written in a compact form by introducing a state vector $\state=(\rho, \rho\vel, \rho e)$:
\be
\ddt{\state} + \int_{V} \grad\cdot\flux dV = \source\label{eq:state},
\ee
where $\flux=(\rho\vel, \rho\vel\vel, (\rho e + P)\vel)$ is the flux function, $\source = (0, -\rho\grad\Phi, -\rho\vel\cdot\grad\Phi$) is the source function \lp{and $V$ is an elemental volume.}

To solve equation (\ref{eq:state}), we adopt the same finite volume strategy as S10.  We refer the interested reader to S10 for a more detailed discussion of the scheme.  Here, we only briefly describe the scheme to document the algorithm we have implemented and to highlight the differences between our scheme and that of S10.

For each cell, the integral over the volume of the $i$th cell defines the charge of the $i$th cell, $\charge_i$, to be
\be
\charge_i = \int_{V_{i}} \state dV = \state_i V_i,
\ee
where $V_i$ is the volume of the cell.
As do S10, we then use Gauss' theorem to convert the volume integral over the divergence of the flux in equation (\ref{eq:state}) to a surface integral
\be
\int_{\partial V_{i}} \grad\cdot\flux dV = \int_{\partial V_{i}} \flux\cdot\normal dA,
\ee
\lp{where $\partial V_{i}$ is the boundary of the cell.} We now take advantage of the fact that the volumes are Voronoi cells with a finite number of neighbours to define an integrated flux
\be
\sum_{j \in {\rm neighbors}} \fluxV_{ij} A_{ij} = \int_i \flux\cdot\normal dA,
\ee
where $\fluxV_{ij}$ and $A_{ij}$ are the average flux and area of the common face between cells $i$ and $j$.
The discrete time evolution of the charges in the system is given by
\be
\charge_i^{n+1} = \charge_i^n + \Delta t \sum_j \hat{\fluxV}_{ij} A_{ij} + \Delta t\sourceV_i, \label{eq:time evolution}
\ee
where $\hat{\fluxV}_{ij}$ is an estimate of the half time-step flux between the initial $\charge_i^n$ and final states $\charge^{n+1}_i$ and $\sourceV_i^{(n+1/2)} = \int_i \source dV$ is the time-averaged integrated source function.

We estimate the flux $\hat{\fluxV}_{ij}$ across each face as follows.
\begin{enumerate}[(i)]
 \item Use the gradient estimates at the initial time-step to predict the half time-step cell centered values.
 \item Drift the cells a half time-step and rebuild the Voronoi tessellation at the half time-step. This step is new compared to C17.
\item Estimate the half time-step state vector (in the rest frame of the moving face) at the face centre (\facer) between the neighboring $i$ and $j$ cells by linear reconstruction. \label{item:half step}
 \item Estimate the (half time-step) velocity $\facev$ of the face, following S10, and boost the state vector from the ``lab'' frame \lp{(the rest frame of the simulation box)} to the rest frame of the face to find the flux along the normal of the face. I.e., in the direction from $i$ to $j$.
 \item Estimate the flux $\hat{\fluxV}_{ij}$ across the face using an HLL or HLLC (or HLLD for MHD, Chang, in preparation) Riemann solver implemented following \citet{toro2009riemann}.
 \item Boost the solved flux back into the ``lab'' frame.
\end{enumerate}
We can then use the estimated fluxes to time-evolve the charges $\charge_i$ following equation (\ref{eq:time evolution}) and using the full time-step $\dt$ and apply changes owing to the source terms.

\subsubsection{Half Time-Step Predictions on a Moving Voronoi Mesh}

We note from the above that we compute the half time-step Voronoi cells to achieve true second-order time integration.  This is a major change from C17, who used the initial time-step cells.  Here we derive the estimate for the time derivative for the half time-step prediction step as this estimate was insufficiently described by C17. 

The Euler equations written in conservative form are equations (\ref{eq:continuity}) to (\ref{eq:energy}).  Following C17, let's assume that the mesh generating point moves with velocity \meshv.  Boosting (in a Galilean sense) into the reference frame of the moving point, equations (\ref{eq:continuity}) to (\ref{eq:energy}) become
\be
\ddt{\rho} + \grad\cdot\rho\velp &=& 0 \label{eq:continuity1},\\
\ddt{\rho\velp} + \grad\cdot\rho\velp\velp + \grad P &=&-\rho\grad\Phi\label{eq:momentum1}
\ee
\begin{flushleft}
and
\end{flushleft}
\be
\ddt{\rho e'} + \grad\cdot\left(\rho e' + P\right)\velp &=& -\rho\velp\cdot\grad\Phi\label{eq:energy1},
\ee
where $\velp = \vel-\meshv$ and $e'= \epsilon + v'^2/2$.  Expanding this out, we can write 
\be
\ddt{\rho} + \grad\cdot\rho\vel - \meshv\cdot\grad\rho &=& 0 \label{eq:continuity2},\\
\ddt{\rho\vel} + \grad\cdot\rho\vel\vel - \meshv\cdot\grad \rho\vel + \grad P &=&-\rho\grad\Phi\label{eq:momentum2}
\ee
\begin{flushleft}
and
\end{flushleft}
\be
\ddt{\rho e} + \grad\cdot\rho e\vel - \meshv\cdot\grad \rho e+ \grad\cdot P\vel &=& -\rho\vel\cdot\grad\Phi\label{eq:energy2}.
\ee
Equations (\ref{eq:continuity2}) to (\ref{eq:energy2}) provide estimates for the time derivatives of the conserved variables, $\rho$, $\rho\vel$ and $\rho e$ with the estimated gradients for the conserved variables calculated as described by C17. This allows us to formally achieve the same second order level of accuracy in time integration as \citet{2016mnras.455.1134p}; \citep[see also][]{2011ApJS..197...15D}, but there are some differences which we discuss below.

\subsection{Multistepping}\label{sec:multistep}

\changaMM\ as described by C17 imposes a universal time-step. Large speedups for problems with large dynamic range are possible with individual time-steps. Hence, to greatly enhance the applicability of \changaMM\ to the problem of CEE, we have implemented an individual time-step scheme for \changaMM.

The basic (universal) time-stepping algorithm for a second order accurate (in time) integrator can be broken down into 3 stages, the initial time (a), the half time-step (b) and the full time-step (c).
\begin{enumerate}[(a)]
 \item \textit{Initial time $t=0$:} determine Voronoi cells using current positions of mesh-generating points. Estimate gradients and construct half time-step predictions. Zero out the changes to the charges, e.g., $\delta\charge=0$.
 \item \textit{Half time-step $t=0.5\dt$:} drift Voronoi cells to half time-step positions. Construct Voronoi cells at half time-steps. Perform linear reconstruction to the half time-step cell faces and compute fluxes.  Perform a Riemann solution and incorporate source terms with the {\it full} time-step. Place these changes into $\delta\charge$.
 \item \textit{Full time-step $t=\dt$: } drift Voronoi cells to full time-step positions. Advance charges to be $\charge^{n+1} = \charge^{n} + \delta\charge$. Reset the state to be the new initial time.
\end{enumerate}

In comparison, \citet{2016mnras.455.1134p} achieve the same formal second order accuracy in two steps, an initial and full time-step.  Here, they make a full time-step prediction for the primitives, perform a Riemann solution using the initial Voronoi cells and faces for a half time-step and perform a closing Riemann solution (again a half time-step) using the final Voronoi cells and faces at the full time-step and the full time-step prediction for the primitives.  The two methods differ in the comparative numbers of Voronoi tessellations and Riemann solutions.  In particular, we only need to solve the Riemann problem once, whereas \citet{2016mnras.455.1134p} require two solutions of the Riemann problem (with associated linear reconstructions).  However, we require an extra Voronoi calculation at the half time-step.  Depending on the relative cost of Riemann solution and the Voronoi calculation -- expensive for tabulated EOSs -- the two schemes have their relative merits.

\begin{figure*}
  \includegraphics[width=0.8\textwidth]{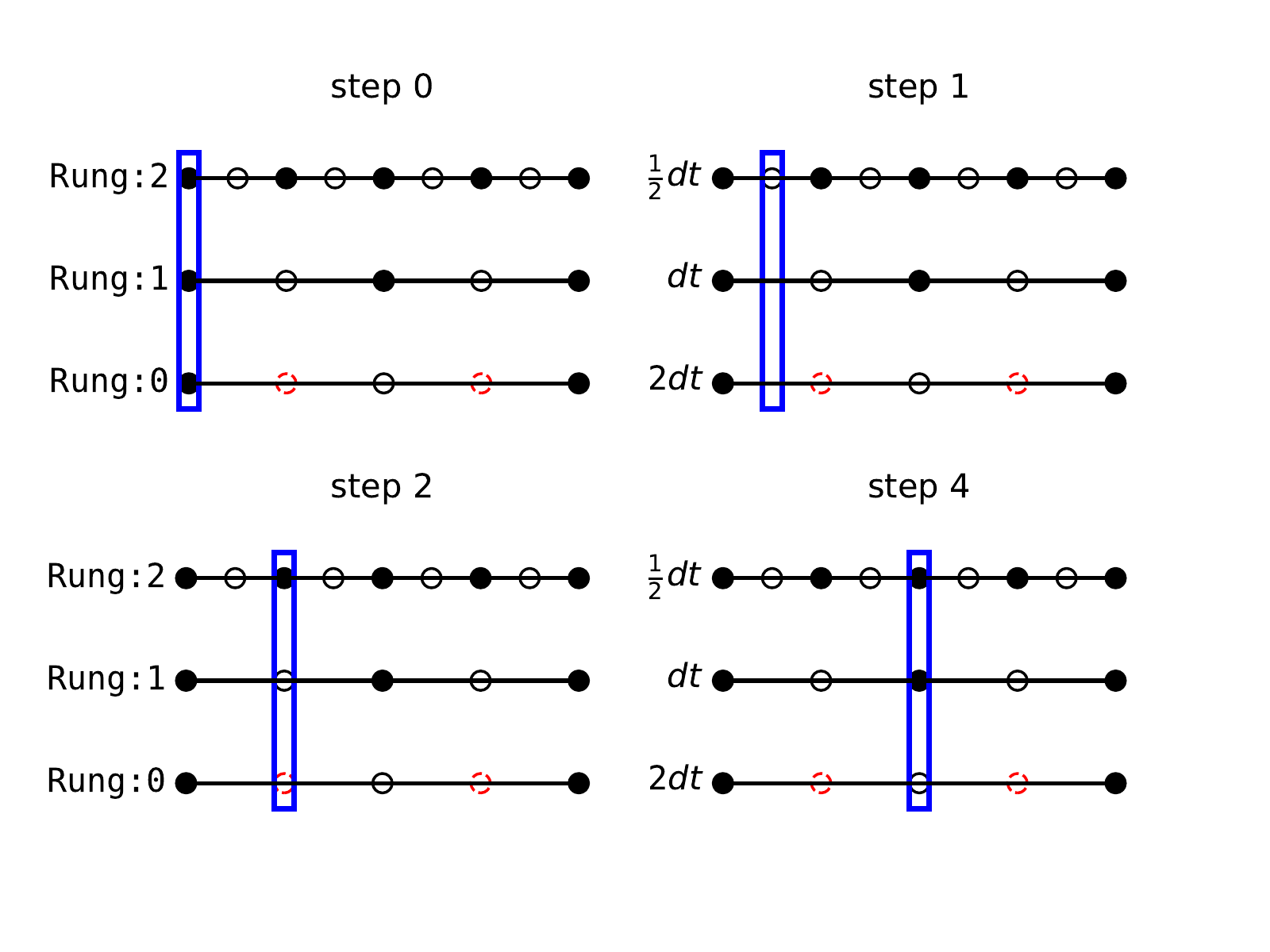}
    \caption{Integration of time-stepping for individual time-steps.  Starting at the lowest rung (rung 0), the time-step is decreased by 2 for each higher rung.  Full and half time-steps at each rung are shown as solid and empty circles, respectively.  The time-stepper (blue rectangle) steps at the smallest half time-step, which in this case is at $0.25\dt$.  All particles drift along at this smallest time-step following the time-stepper. Half time-step procedures (full time-step Riemann solutions on half time-step Voronoi cells) are executed only at half time-steps relevant to the particular rung level.  Full time-step procedures (accumulating all $\delta\charge$'s and full time-step gradients and half time-step prediction) are executed at full time-steps.  Note that only one half time-step at one rung is relevant for each step of the time-stepper, but multiple full time-steps at more than one rung can be executed at the time-stepper's current position. \label{fig:timeline}}
 \end{figure*}

To adapt our scheme for multistepping, we note that the full time-step for any cell involves actions at a full time-step (a,c) and the half time-step (b).  Let us then consider the time advancement of three separate levels or rungs at 0,1,2, where the time-step decreases by a factor of 2 at every level or rung in Fig. \ref{fig:timeline}. In the interest of keeping terminology between the code and this paper synchronized, we drop the term level and refer to the different time-step levels as rungs.

In Fig. \ref{fig:timeline}, the full time-steps are represented by filled circles, while the half time-steps are represented by empty circles.  At step~0, all three rungs are synchronized at their respective full time-steps: all the circles are black and step (a) is executed on all three rungs.  The particles are then drifted $0.25 \dt$ to step 1. The half time-step at rung 2, and the half time-step reconstruction and Riemann solution is performed for rung 2.  Note that the Riemann solution is for the full time-step $0.5 \dt$ at rung 2 so that when the $\delta\charge$ is summed at the next step, the $\charge$s for rung 2 are updated to the full time-step.  Now at step 2, the full time-steps (c) and initial time-steps (a) are performed at rung 2 and the half time-step (b) is performed for rung 1.  In addition, cells at rung 0 that share a face with a cell at rung 1 (denoted by red-dashed circles) also have a Riemann solution at the common face at the rung~1 half time-step.
We skip step 3 as it is the same as step 1 and now proceed to step 4. Here, we do the full time-steps for rungs 1 and 2 and the half time-step for rung 0.

From this discussion, we can conclude that for the half time-step for rung $i$, full time-steps are executed from all rungs $j > i$.  Conversely, there can only be one half time-step for only one rung per (drift) time-step. This is the half time-step of the highest rung. In addition, cells only switch rungs during their full time-steps to ensure consistency of the second-order time integration and they can only reduce their rung when their new reduced rung is at its full time-step.

One additional feature that we have included in our implementations is the smoothing of time-steps by large propagating velocity fluctuations, or shocks.  This is a well known issue in SPH, for which \citet{2009ApJ...697L..99S} show that large localized changes in the time-step can lead to incorrect results.  They also showed that smoothing out time-stepping changes over a SPH kernel such that the time-step variation is limited to a factor of 4 mitigates these issues.  We have adopted a similar approach here in that we limit the cell's time-step to be less than $\sqrt{2}$ of the minimum time-step of its neighbours.  This results in time-step changes of a factor of two occurring over two cells away from a minimum cell.  We find that this smoothing allows for stable integrations in \changaMM.

This multistepping algorithm was applied to a simple CEE simulation to test the speedup factor. At each step, we find speedups ranging from 3 to 9 times that of the universal time-stepping algorithm, resulting in an average speedup of 4 to 5 times.

\subsection{Realistic Equation of State}\label{sec:eos}

In our original implementation of \changaMM\ described by C17, we only used an ideal, adiabatic equation of state.  However, it is straightforward to extend \changaMM\ to arbitrary equations of state.  For CEE, we have elected to use the equations of state implemented in the open source stellar evolution code, \mesa\ \citep{2011ApJS..192....3P,2013ApJS..208....4P,2015ApJS..220...15P}.  The \mesa\ equation of state relies on several equations of state relevant over the different regimes of stellar structure, allowing \mesa\ to span the full range from giant planets and white dwarfs to massive stars \citep{2011ApJS..192....3P}. 

For a hydrodynamic solver, the equation of state can be called many times -- during the update of cell centred variables, during the reconstruction step, during the flux calculation and during the Riemann solution (to determine sound speeds).  We have found that tabulated equations of state are much too slow to be called so many times during a single solve.  In particular, an average Voronoi cell has about 20 faces over which a Riemann solver must be executed. This means at least 20 calls to the equation of state to find the sound speed (for linear or higher order reconstruction).  Instead, we only make one call per step in \changaMM\ -- during the update of cell centred variables.  To find values of these variables on the faces, we estimate the relevant thermodynamic quantities using linear reconstruction of the density, energy and adiabatic index.  We have found that this results in a factor of 25 speedup in the Riemann solver.


\subsection{Initial Conditions}\label{sec:ICs}

We now discuss the development of appropriate initial conditions for \changaMM.  As stated in the Introduction, \citet{2016ApJ...816L...9O} previously performed a similar moving-mesh simulation of CEE.  Here, we adopt similar initial conditions.  We first use \mesa\ to evolve a 2 \msun\ star \lp{with metallicity $Z=0.02$} from the pre-main sequence to the red giant phase.  We stop when the star reaches 52 \rsun\ with a He core mass of 0.36 \msun.  Compared with that of \citet{2016ApJ...816L...9O} our red giant is slightly larger and its He core mass is slightly smaller.  This \lp{\sout{is likely} could be} due to the updates in the \mesa\ code since their original result \lp{or discrepancies in our \mesa\ inlists}.  From the \mesa\ output, we take the entropy and hydrogen fraction.  For the core, we take the total mass at a density that is $50$ times greater than the mean density of the red giant, giving a core mass $M_{\rm{c}} = 0.379 \msun$.  This corresponds to a core radius $R_{\rm{c}} = 1.99 \rsun$, which we use as the core gravitational softening length.  Because of the great difference in density between the He core and the H envelope, we model the core as a dark matter particle with mass $M_{\rm{c}}$ and softening radius of $R_{\rm{c}}$ similar to the initialization of \citet{2016ApJ...816L...9O}'s simulations. We then take the entropy profile and construct a star of mass $M-M_{\rm{c}}$, with an entropy profile which matches that of the original star and contains a dark matter particle core.  This yields a radial profile of density, temperature and H-fraction that can be mapped to a particle (mesh generating point) profile.

We construct an appropriate particle mesh for the star from a precomputed glass distribution \lp{of points embedded in a 3-D cube. \sout{that has been periodically replicated} We periodically replicate this glass distribution} to produce sufficient numbers of particles.  We assume that each particle is of equal mass and rescale them to the appropriate radial position based on the computed $M(r)$ from \mesa.  These particles are also endowed with the radially interpolated temperature and H-fraction.  The total number of particles representing the star is $3\times 10^5$, which is smaller than the $2\times 10^6$ particles used by \citet{2016ApJ...816L...9O}.  Outside of the star, we include a low density atmosphere of $10^{-13}$ g cm$^{-3}$ with temperature $10^{5}$ K that extends out to the total box size of $3.5\times 10^{14}$ cm ($5000~\rsun$), with periodic boundary conditions at its edges.  The total number of particles in the simulation is $8\times 10^5$.

To lower the computational cost, we use a mesh refinement algorithm to decrease the number of gas particles in the atmosphere far from the star. We define a scale factor $\mathcal{S}(r) = (r/R_{*})^{n}$ where $R_{*}$ is the radius of the star, $r$ is the spherical radius and $n$ is an adjustable parameter which we have set to $n = 2/3$ in this case. Starting with the same uniform glass distribution as for the star, the linear spacing between particles is increased by $\mathcal{S}$ and their mass is increased by $\mathcal{S}^{3}$, preserving the external density.

The profiles produced from the mapping from 1-D stellar evolution models to fully 3-D hydrodynamics simulations are not in perfect hydrostatic balance owing to discretization errors.  Previously C17 used this fact to compare (roughly) hydrostatic stars experiencing oscillations using the MM and SPH solver.  Here we are interested in the equilibrium solution in the MM solvers.  Thus, following the mapping from 1-D to 3-D, we damp spurious velocities and energies and allow the star to relax to a final configuration.  We note that this is the same procedure that we have used previously for merging white dwarfs \citep{zhu+12} and was also used by \citet{2016ApJ...816L...9O} for giant star initial conditions.

\begin{figure}
  \includegraphics[width=0.5\textwidth]{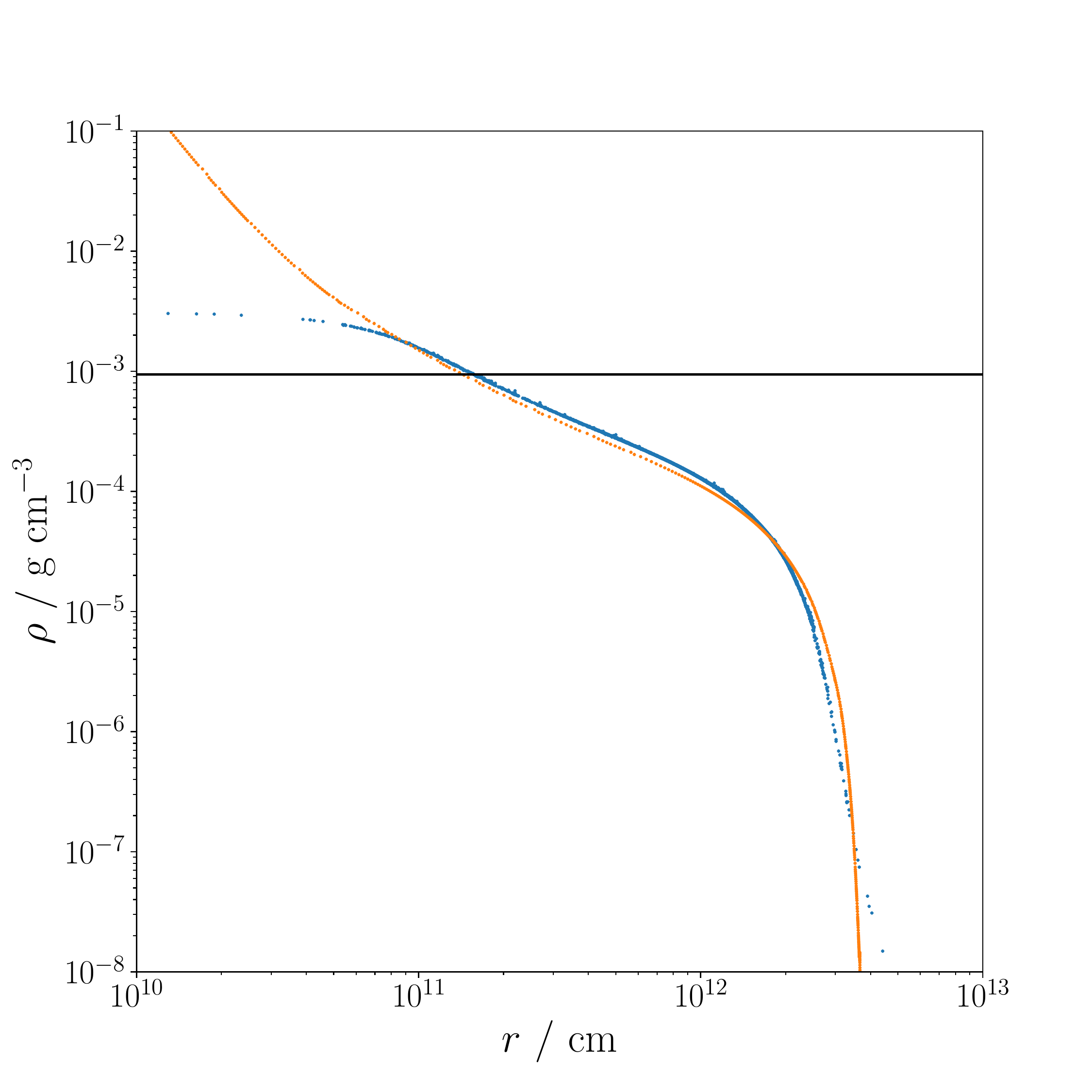}
    \caption{Relaxed \lp{density $\rho$ against radius $r$} profile from \changaMM\ (blue circles) compared to the profile from \mesa\ (orange circles). The horizontal line denotes the point where the density in the \mesa\ profile is 50 times greater than the mean density.  \label{fig:radprof}}
\end{figure}

We show the result of this relaxation in Fig. \ref{fig:radprof} as blue circles and compare this to the radial profile from \mesa\ (orange circles).  Here we downsample the number of points from \changaMM\ by a factor of 100.  The horizontal line denotes the point where the density in the \mesa\ profile is 50 times greater than the mean density.  The mass inside of this radius is represented by a dark matter particle with a gravitational softening length equal to the radius.  The star is then recomputed to follow the same entropy--mass relation (corrected for the inclusion of the dark matter particle) and then relaxed in \changaMM.  The envelope follows the original \mesa\ profile quite faithfully in spite of the differences in the physical structure.  The inner profile approaches a fixed density asymptotically owing to the softened gravitational forces from the dark matter core.

We represent the 1 \msun\ companion as a dark matter particle that is in an initially circular Keplerian orbit at the red giant radius \lp{\sout{e.g., initial semi-major axis}} $a=52\rsun$. This follows the initial conditions of \citet{2016ApJ...816L...9O} to facilitate a direct comparison. Although this neglects the evolution of the binary prior to CEE, we compensate by altering the dynamics of the giant. During the phase leading up to CEE, tidal forces from the companion on the envelope are expected to spin up the rotation of the envelope \citep{1996ApJ...460L..53S} to a significant fraction of its breakup velocity. We have implemented this corotation between the envelope rotation and orbital motion into our simulations by following the scheme of \citet{2018ApJ...863....5M}. Within the envelope, we assume rigid body rotation and initialize the velocity as

\be
v_{\phi} = f_{\rm{cr}} \Omega_{\rm{orb}} R_{\rm{cyl}}, \label{eq:v_in}
\ee
\begin{flushleft}
but give the atmosphere a velocity
\end{flushleft}
\be
v_{\phi} = \frac{ f_{\rm{cr}} \Omega_{\rm{orb}} R_{*}^{2} \sin^{2}(\theta) }{ R_{\rm{cyl}} }. \label{eq:v_out}
\ee

\begin{flushleft}
Here, $f_{\rm{cr}}$ is an adjustable parameter, $\phi$ is the azimuthal angle, $\theta$ is the polar angle, $\Omega_{\rm{orb}}$ is the orbital frequency of the red giant and companion, $R_{*}$ is the radius of the giant and $R_{\rm{cyl}}$ is the cylindrical radius from the rotation axis of the giant. We note that (\ref{eq:v_in}) and (\ref{eq:v_out}) ensure that the velocity is continuous at the surface of the giant.
\end{flushleft}

Here, we run two simulations in which the giant is in different states of corotation, 95 per cent corotation ($f_{cr} = 0.95$) and 0 per cent corotation ($f_{cr} = 0$). We note that the 95 per cent corotation case follows the simulations of \citet{2016ApJ...816L...9O} and \citet{2012ApJ...746...74R}. \lp{The choice to set $f_{cr}$ to less than unity is also motivated by \citet{2018ApJ...863....5M} who found that orbital angular momentum lost by the companion prior to the onset of a common envelope phase does not necessarily go into spinning up the envelope, leading to a desynchronization between the orbital frequencies of the companion and envelope.}

\section{Results}\label{sec:results}

\begin{figure*}
  \includegraphics[width=1.0\textwidth]{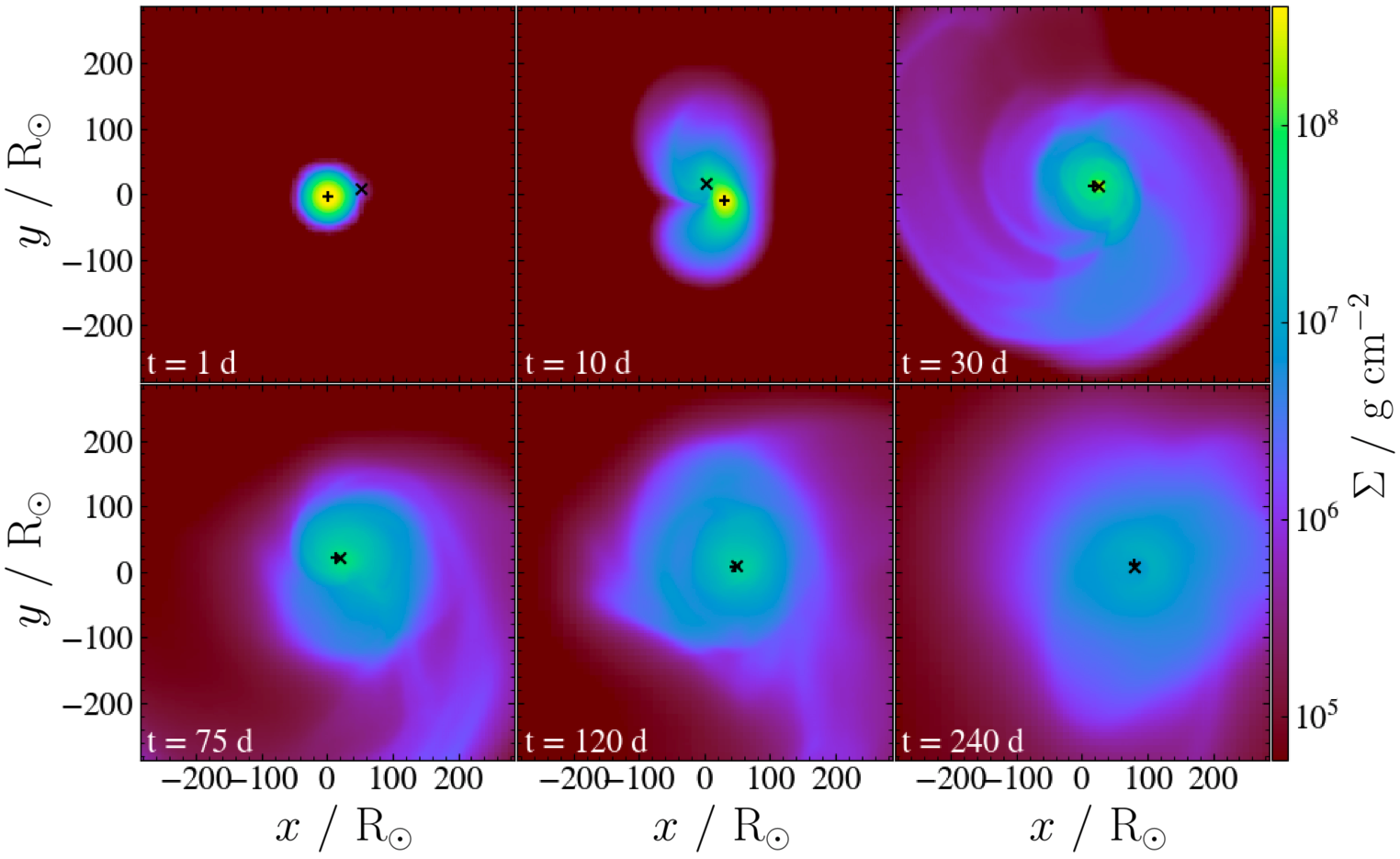}
    \caption{Density projections $\Sigma$ with 95 per cent corotation. The + sign marks the red giant core and the $\times$ marks the companion.
    \label{fig:corotframes}}
\end{figure*}

\begin{figure*}
  \includegraphics[width=1.0\textwidth]{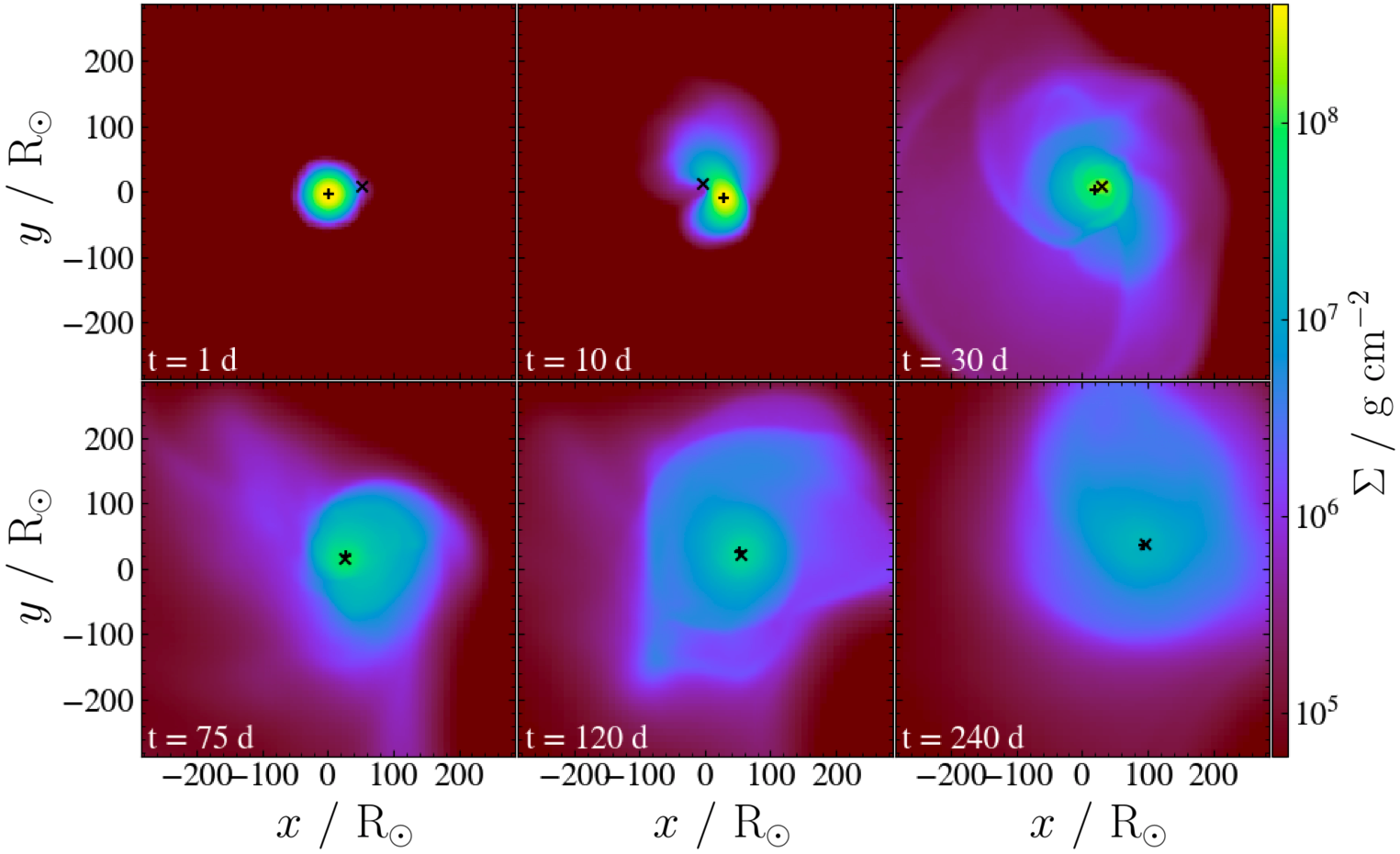}
    \caption{Density projections $\Sigma$ with 0 per cent corotation. The + sign marks the red giant core and the $\times$ marks the companion.
    \label{fig:mesaframes}}
\end{figure*}

Starting with the above initial conditions, we simulate the binary for 240 d and show several density projections at $t = $ 1, 10, 30, 75, 120 and 240 d in Figs. \ref{fig:corotframes} and \ref{fig:mesaframes}. \lp{We note that the simulation period is about 10 times the initial orbital period of 25 d.}

In both simulations, the companion experiences an initial plunge into the envelope of the giant that lasts about 15 days, throwing off a large tidal tail (Figs. \ref{fig:corotframes} and \ref{fig:mesaframes} upper centre) and greatly decreasing the separation between the companion and core. The companions then continue to spiral in as their orbital energy is transferred to the gas; \lp{\sout{as shown in Fig. \ref{fig:mech}}} the spiral shocks facilitating this transfer can be seen in the projections (Figs. \ref{fig:corotframes} and \ref{fig:mesaframes} upper right). We see smaller tidal tails thrown off at $t=75$ d (lower left) and $t=120$ d (lower centre), as well as several others. The simulation ends well before the outflow reaches the edge of the simulation box.

\subsection{Envelope Ejection}\label{sec:unbound}

\begin{figure}
\centering
    {
    \includegraphics[width=0.5\textwidth]{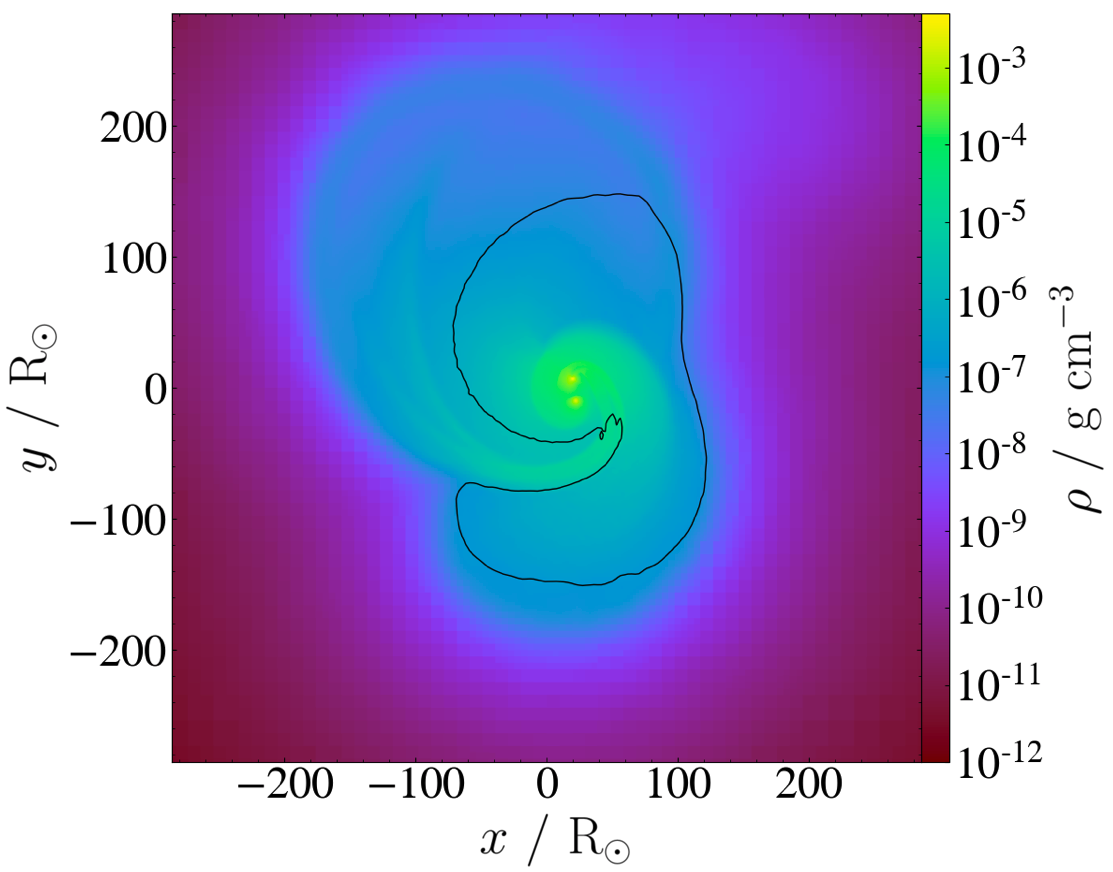}
    }
    \caption{A density slice through the $z=0$ plane at $t=15$ d of the 95 per cent corotation simulation. The black boundary encloses the gas that is bound to the binary, revealing an unbound tidal tail being ejected.}
    \label{fig:slices}
\end{figure}

The total energy of each gas particle is given by
\be
\Etoti = m_{i} [ \frac{1}{2} (\vel_{i} - \velCM)^{2} + \phi_{i} + I_{e,i} ] \label{eq:Etot}
\ee
\citep{2015MNRAS.450L..39N}, where $m_{i}$, $\vel_{i}$, $\velCM$, $\phi_{i}$ and $I_{e,i}$ are the mass, velocity, centre of mass velocity of the bound material, gravitational potential and specific internal energy of each particle. Gas particles with a negative total energy are bound to the binary, while those with a positive total energy are unbound. Fig. \ref{fig:slices} shows a density slice in the $z=0$ plane at $t=15$ d in the 0 per cent case. The black contour encloses the matter bound to the binary.

The kinetic energy is computed relative to the velocity of the centre of mass (CM) of the \textit{bound} matter; that is, the bound gas as well as both dark matter particles. However, because the total energy is needed in order to determine which gas particles are bound, we used an iterative scheme to find the velocity of the CM. \\

\begin{enumerate}[(i)]

\item The CM velocity is assumed to be zero. \\

\item The velocities of the gas particles are calculated relative to the CM velocity. \\

\item These velocities are used to find the kinetic energies of the gas particles, which are used to determine which particles are bound and which are not. \\

\item A new CM velocity $\velCM$ is found for everything \textit{except} the unbound particles,

\be
\velCM = \frac{ \mcore\vcore + \mcomp\vcomp + \mbound \vbound }{ \mcore + \mcomp + \mbound },
\ee
where $\mcore$ and $\mcomp$ denote the dark matter core and companion, respectively, $\mbound$ denotes the bound gas, \mcore\vcore, \mcomp\vcomp, and \mbound\vbound\ and \vcore, \vcomp, and \vbound\ are the momenta and velocities of the dark matter core and companion and bound gas, respectively. \\

\item Steps (ii) to (iv) are repeated until $\velCM$ reaches convergence.

\end{enumerate}

\begin{figure}
  \includegraphics[width=0.5\textwidth]{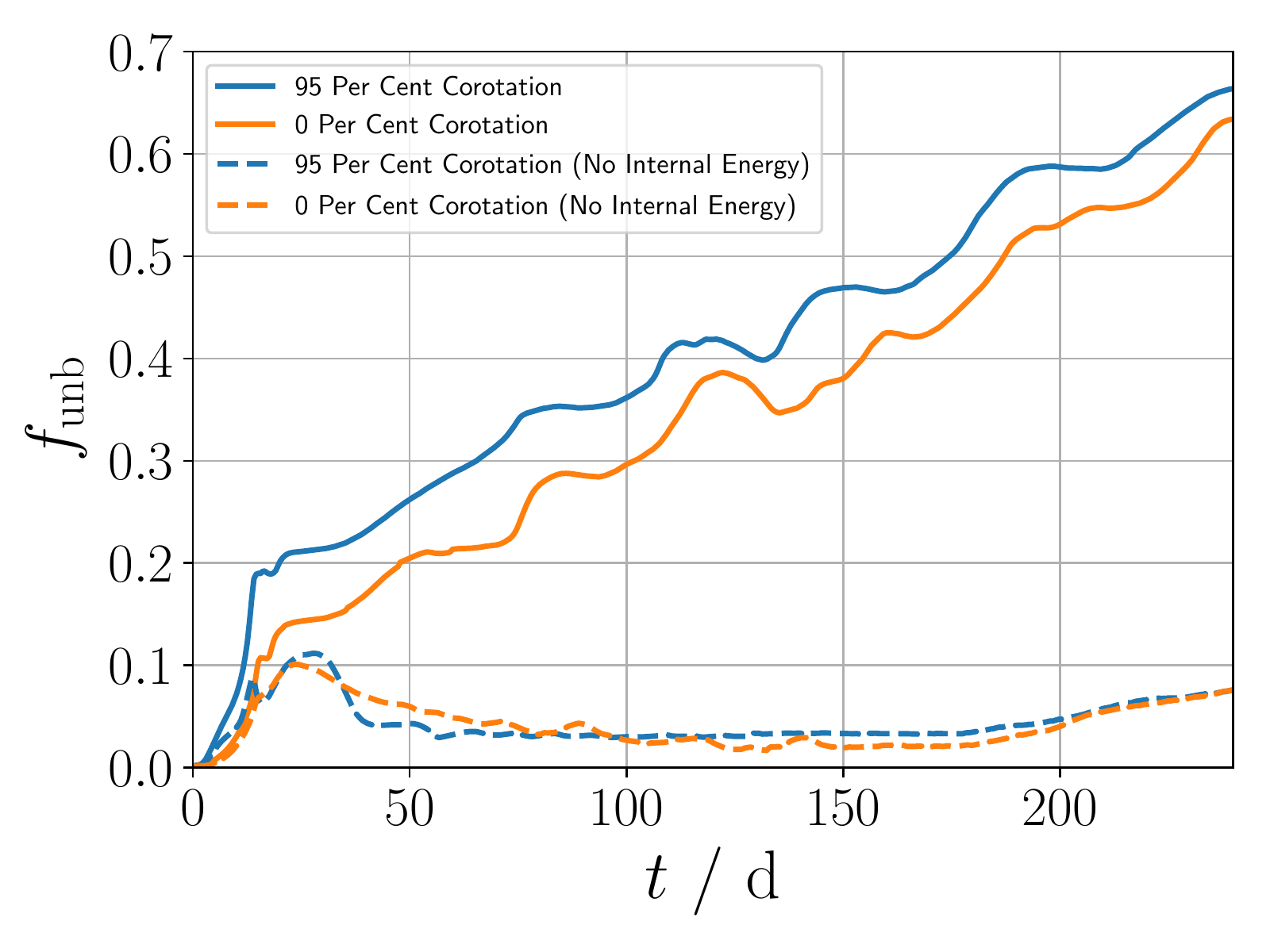}
    \caption{The fraction of mass of the envelope that has acquired enough energy to be unbound from the system. The large tidal tail thrown off in the initial plunge can be seen from 0 to 15 d, with several subsequent tails. \lp{A calculation of $f_{\rm unb}$ neglecting the internal energies is also shown (dashed lines).}  \label{fig:unbound}}
\end{figure}

This is similar to the method of \citet{2013ApJ...767...25G}. With the total energies of all particles known, we can find the mass of the unbound gas as a fraction of the total mass of the envelope $m_{\rm env}$,
\be
\funbound = \frac{ \munbound }{ m_{\rm env} },
\ee
\pc{which we refer to as the ejection efficiency.}
This is shown in Fig. \ref{fig:unbound}. \pc{\sout{We refer to this as the ejection efficiency.}} Qualitatively, the simulations are very similar in this regard. They show a large ejection of gas during the initial plunge as well as during the ejection of additional tidal tails. Unsurprisingly, the system set into corotation consistently exhibits a higher ejection efficiency owing to the inclusion of additional energy and angular momentum. We find that matter continues to be ejected up until the end of the simulation period, ending with an efficiency of 66 per cent with corotation and 63 per cent for no corotation.

\lp{We also show in Fig. \ref{fig:unbound} the ejection efficiency that would be obtained if the internal energies were neglected. We find that only 8 per cent of the envelope is unbound in both cases, which is similar to the reported ejection efficiency in \citet{2016ApJ...816L...9O}.} \pc{Such a situation can be reached if the envelope expands sufficiently such that it loses energy rapidly by radiation.}  \lp{This indicates that the internal energy plays a large role in unbinding the envelope.} \pc{It also suggests that further work on CEE may not be able to ignore radiative cooling because of its possible large effects.}

%

\subsection{Orbital Parameters}\label{sec:orbital}

\begin{figure}
  \includegraphics[width=0.5\textwidth]{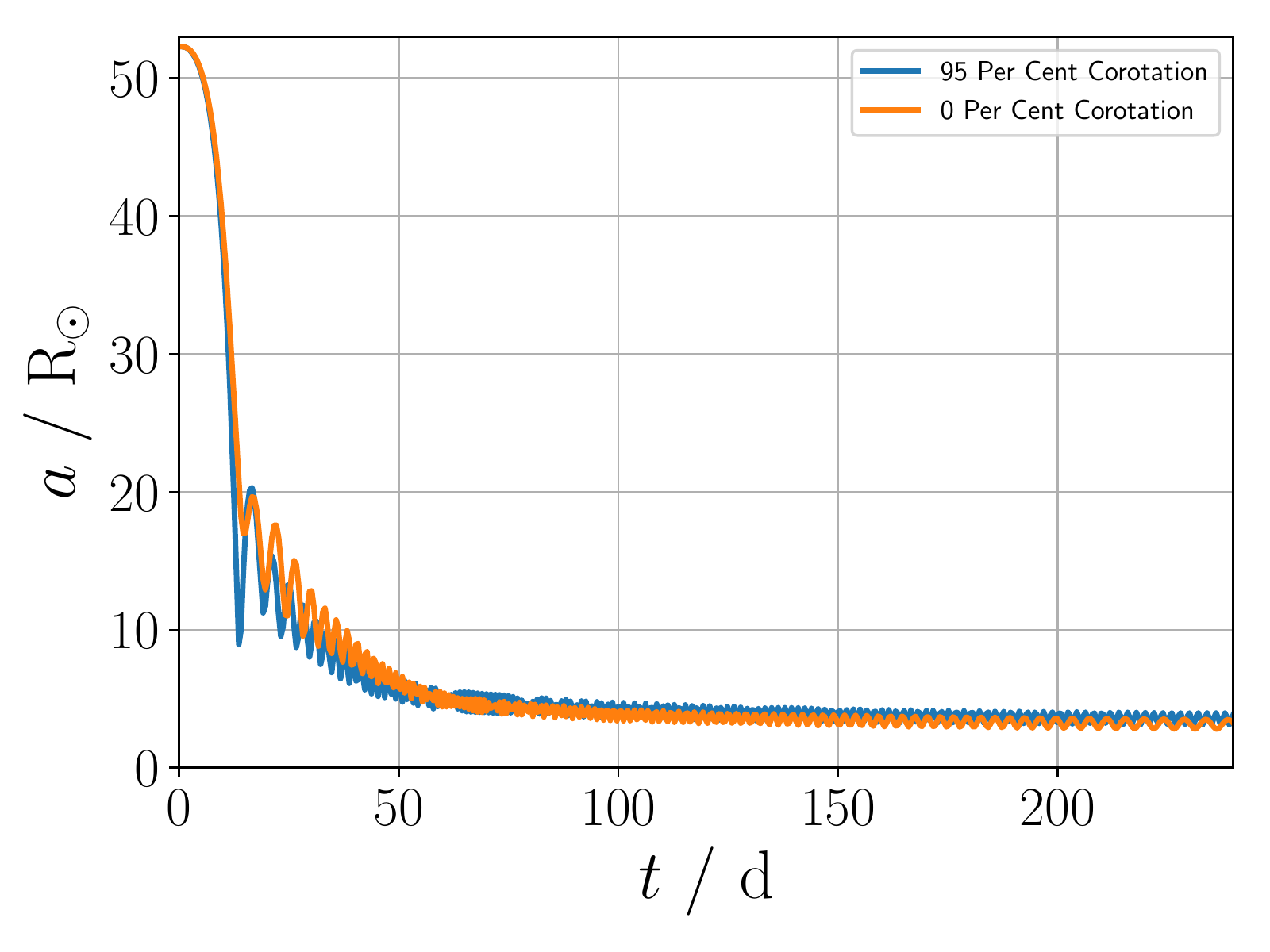}
  \includegraphics[width=0.5\textwidth]{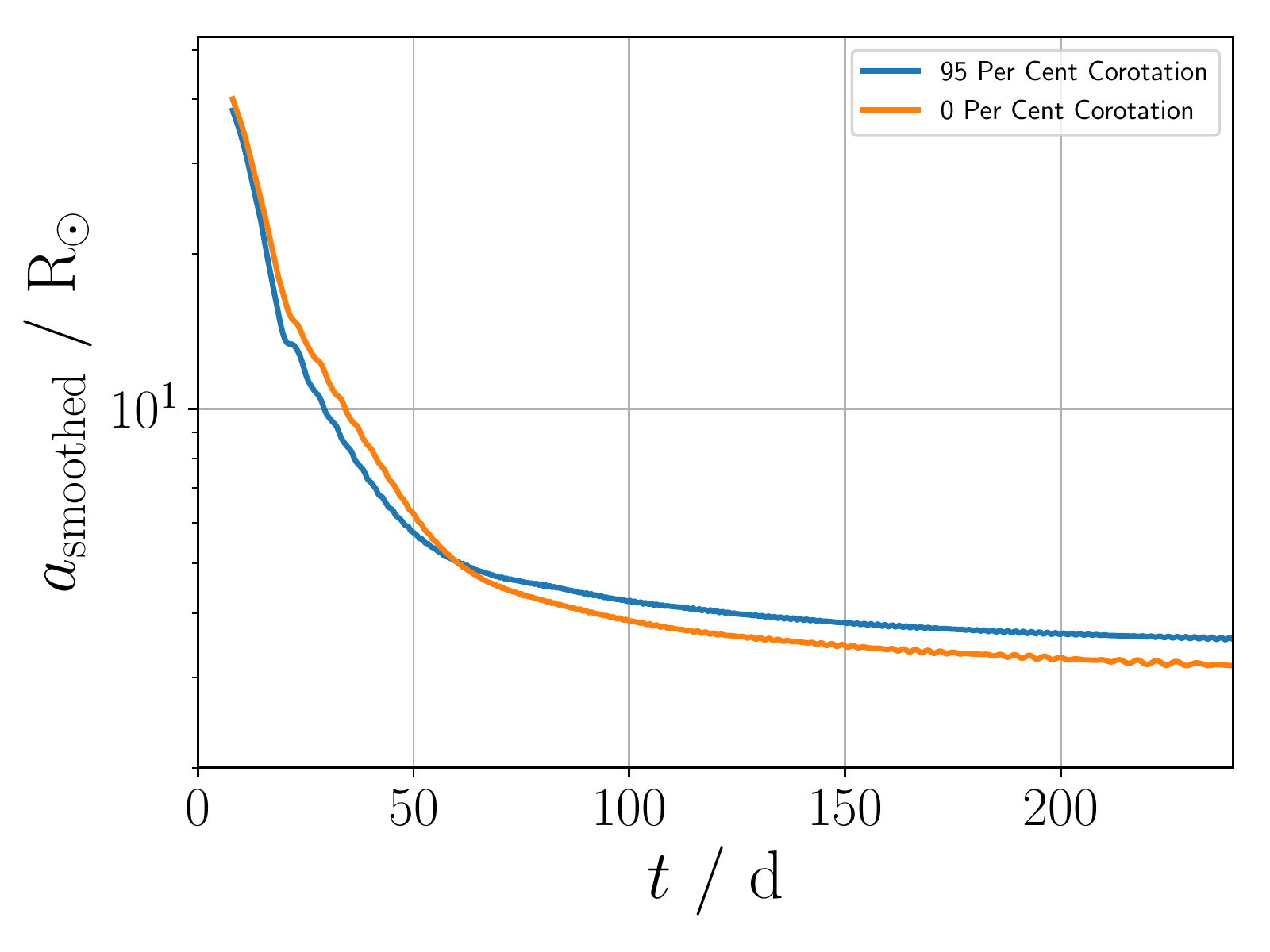}
    \caption{\textit{(Top)} Separation $a$ between the stellar core and companion in each case.
    \textit{(Bottom)} The separations smoothed over time are shown on a logarithmic scale for ease of reading. Here we can see that the simulation with no corotation results in a smaller separation by 0.4 $R_{\odot}$.
    \label{fig:separation}}
\end{figure}

The separations between the stellar core and companion are shown in Fig. \ref{fig:separation}. For ease of reading, we also show the separations on a logarithmic scale and smooth them over a period of 15 d. That is, each separation $r_{i}$ is taken to be the average of all separations within an interval of 15 d centred on $r_{i}$.

After one orbit, both binaries have reduced their orbital separation to less than half the initial separation, and they continue to spiral in at a slower rate. Although the simulation with corotation initially falls to a smaller separation, it is eventually surpassed by its counterpart. The final (smoothed) separations are 3.6 $\rsun$ with corotation and 3.2 $\rsun$ with no corotation.

The eccentricity $e$ can be found from the periapsis and apoapsis distances,
\be
e = \frac{ \rapo - \rperi }{ \rapo + \rperi },
\ee
where \rapo\ and \rperi\ denote the distances of closest and furthest approach. Here \rapo\ and \rperi\ are determined by interpolating the maxima and minima of the separation. The alert reader will notice that this definition is meaningful in this situation because the potential is not Keplerian owing to the distribution and nonaxisymmetry of the unbound gas.  However, this method allows the determination of the eccentricity from only the orbital separations. Both binaries circularize their orbits over time, seeing drops in eccentricity from $e \approx 0.2$ down to $e = 0.1$ by the end of the simulation period.

\begin{figure}
    \includegraphics[width=0.5\textwidth]{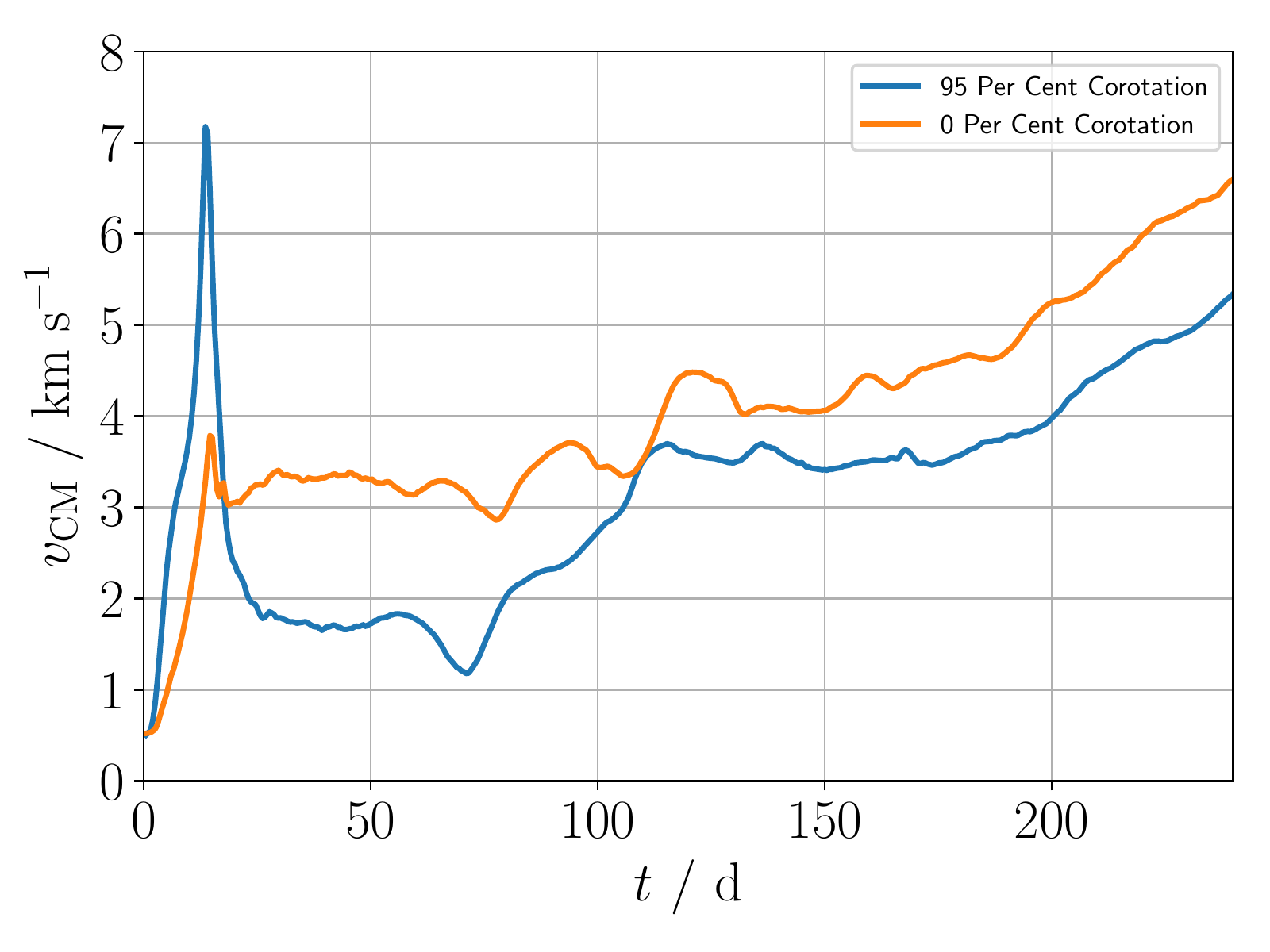}
    \caption{The velocity of the centre of mass of the system $v_{\rm CM}$ over the simulation period. The simulation including corotation feels a large kick during the initial plunge but subsequent kicks quickly slow it to about 1 km s$^{-1}$ less than its non-corotating counterpart.}
    \label{fig:CMvel}
\end{figure}

In addition, each binary receives a kick velocity owing to the ejection of gas from the system, as shown in Fig.~\ref{fig:CMvel}. Interestingly, although corotation results in a larger mass ejection and a large initial spike in CM velocity during the plunge, it quickly drops below the non-rotating case by about 1 km s$^{-1}$.

\section{Discussion and Conclusions}\label{sec:discussion}

In this initial work, we describe the methodology for simulations of CEE using the moving-mesh code \changaMM.  First, we detail a new multistep scheme to speed up problems with large dynamic range.  In the case of CEE, we find speedups of up to 4 to 5 times over our previous single-stepping scheme.  We then implement a realistic equation of state using the libraries in the stellar evolution code \mesa\ and discuss the construction of realistic initial conditions. 

Using these initial conditions, we run two simulations of a 2 \msun\ red giant with a 1 \msun\ companion (thought to be a main sequence star), where we take the rotation of the envelope to be either 95 or 0 per cent of the (initial) orbital corotation.  After 240 d, we find that the orbital separation of the red giant core and the companion has been reduced to 3.6 \rsun\ and 3.2 \rsun, respectively.  This is a reduction in the separation by a factor of 15. Most of the envelope is ejected (66 and 63 per cent respectively).  \lp{\sout{We argue that the majority of the energy required to eject the envelope comes from the orbital energy of the core and companion.}}  


Our envelope ejection efficiencies are much larger than many of the results of previous work.  However, three-dimensional numerical computations of CEE have found a wide variance in the ejection efficiency of the envelope. Early work \citep[for example,][]{1987PhDT.......113D,1988ApJ...329..764L,1994ApJ...422..729T,1998ApJ...500..909S,2000ApJ...533..984S,1995ApJ...445..367T,1996ApJ...471..366R} on 3-D numerical simulations of CEE found envelope ejection efficiencies from about 10 to 80 per cent.  However, the resolutions of these early simulations were poor. More recent work \citet{2012ApJ...744...52P,2012ApJ...746...74R,2015MNRAS.450L..39N,2016ApJ...816L...9O} also find a large variance in the efficiencies (about 5 to 100 per cent). For instance, \citet{2012ApJ...744...52P} found that only 15 per cent of the mass was unbound in their simulations.  \citet{2012ApJ...746...74R} found that about 25 per cent of the mass is unbound, but the unbound mass is still rising at the end of their simulations (see their Fig. 9).  \citet{2015MNRAS.450L..39N} found that between 50 and 100 per cent of the envelope mass is ejected in their simulations, depending on the inclusion of recombination energy. \citet{2016ApJ...816L...9O} found that about 8 per cent of the envelope mass is ejected.  Our results are on the higher side and are similar to what \citet{2015MNRAS.450L..39N} find.  One difficulty in comparing these results is that the initial conditions and simulation parameters are not similar in these different groups.  In this paper, we have adopted initial conditions and parameters similar to those used by \citet{2016ApJ...816L...9O} but our results are dramatically different. The difference may be attributed to different simulation parameters such as the gravitational softening, the parameterization of the dark matter particle that represents the core of the red giant and the companion, or our inclusion of the recombination energy though the use of the \mesa\ EOS.  \lp{Indeed, we have found that the internal energy has a large impact on the ejection efficiency.} \pc{This suggests that the incorporation of radiation physics may be an important next step.}  

It has \pc{also} been suggested that inclusion of recombination energy would result in more efficient expulsion of the envelope \citep[see the review by][]{2017arXiv170607580I}.  
We also note that this point is controversial, as Soker and collaborators have claimed in a series of papers \citep{2017MNRAS.472.4361S,2017arXiv170603720S,2018arXiv180508543S,2018MNRAS.478.1818G} that the energy from recombination is easily transported away either by radiation or convection. \lp{However,} \citet{2018ApJ...858L..24I} \lp{reached the opposite conclusion, that the fraction of recombination energy transported away from the regions in which recombination takes place is negligible.} \pc{The importance of radiative cooling in this case highlights the need for a more careful accounting of radiation effects in future calculations.}

Our work here reinforces the results of \citet{2015MNRAS.450L..39N} which suggests that the entire envelope of the red giant might be ejected during the plunge phase of CEE or at least can be numerically simulated.  Thus, the effects or relevance of the other phases of CEE such as the self-regulated spiral-in phase is unclear.  It may be the case that these phases are important or relevant for different system parameters, though this requires a more detailed exploration of parameter space. 



Finally, we note here that other processes may also aid the expulsion of the envelope.  Much recent work have been focused on energy and momentum injection from jets produced by accretion onto the companion \citep{2015MNRAS.449..288P,2017MNRAS.470.2929M,2018MNRAS.477.2584S,2018MNRAS.tmp.1849C}.  In addition, the transition from laminar flow to accretion flows around the companion star can be complicated and also provide another mechanism by which orbital energy is dissipated \citep{2015ApJ...798L..19M,2015ApJ...803...41M,2017ApJ...838...56M}.  Much of the work have focused on Bondi-Littleton-Hoyle type accretion around a point mass but a typical accretor has a physical scale of a solar radius, which gives it a nontrivial cross section in a red giant envelope.  Typically, we can show that a 1 $R_{\odot}$ accretor encounters 10 to 50 per cent of the envelope in one orbit. This hydrodynamic interaction may play a significant role in the evolution of the common envelope and the binary orbit.  Finally, we mention that it is also suggested that dust formation during CEE can drive winds to expel the envelope \citep{1994MNRAS.270..774S,2018MNRAS.478L..12G}.


\section*{Acknowledgements}

We acknowledge useful conversations with Thomas Quinn and Volker Springel during the development of the multistepping algorithm and refinement on the half time-step prediction.  We also are grateful to Bill Paxton for tracking down a \mesa\ bug in release 10398, supplying a workaround, and correcting it in future versions of \mesa.   We thank S. Ohlmann and R. Pakmor for useful discussions. \pc{We thank the reviewer, Christopher Tout, for a useful and constructive referee report.} We are supported by the NASA ATP program through NASA grants NNH17ZDA001N-ATP, NSF CAREER grant AST-1255469, and the Simons Foundation.
We also used the Extreme Science and Engineering Discovery
Environment (XSEDE), which is supported by National
Science Foundation (NSF) grant No. ACI-1053575.
We also acknowledge the Texas Advanced Computing Center (TACC) at The University of Texas
at Austin for providing HPC resources that have contributed to the research results reported
within this paper (URL: \url{http://www.tacc.utexas.edu}). We also use the {\sevensize YT} software platform for the analysis of the data and generation of plots in this work \citep{yt}.




\bibliographystyle{mnras}
\bibliography{references}


\bsp	
\label{lastpage}
\end{document}